\documentclass[a4paper]{article}

\usepackage{INTERSPEECH2020}
\usepackage{multirow}
\usepackage{url}
\title{Unsupervised Acoustic Unit Representation Learning for Voice Conversion
using {WaveNet} Auto-encoders}
\name{Mingjie Chen, Thomas Hain}
\address{
  Department of Computer Science, University of Sheffield, Sheffield, UK
}
\email{\{mchen33, t.hain\}@sheffield.ac.uk}

\begin{document}

\maketitle

\begin{abstract}
Unsupervised representation learning of speech has been of keen interest in recent years, 
which is for example evident in the wide interest of the ZeroSpeech challenges. 
This work presents a new method for learning frame level representations based on WaveNet auto-encoders. Of particular interest in the ZeroSpeech Challenge 2019 were models with discrete latent variable such as the Vector Quantized Variational Auto-Encoder (VQVAE). However these models
generate speech with relatively poor quality. In this work we aim to address this with two approaches: first WaveNet is used as the decoder and to generate waveform data directly from the latent representation; second, the low complexity of latent representations is improved with two alternative disentanglement learning methods, namely instance normalization and sliced vector quantization. The method was developed and tested in the context of the recent ZeroSpeech challenge 2020. The system output submitted to the challenge obtained the top position for naturalness (Mean Opinion Score 4.06), top position for intelligibility (Character Error Rate 0.15), and third position for the quality of the representation (ABX test score 12.5). These and further analysis in this paper illustrates that quality of the converted speech and the acoustic units representation can be well balanced.   


\end{abstract}

\section{Introduction}
Unsupervised speech representation learning has gained interest of researchers. 
It has been shown that representation learning benefits downstream speech applications such as: speech recognition \cite{Baevski2020vq-wav2vec}, speaker verification \cite{Chung2019} and voice conversion \cite{hsu2017unsupervised}. 
In some recent voice conversion systems \cite{hsu2017unsupervised,yingzhen2018disentangled,chou2018multi}, auto-encoder based models with disentanglement learning objective functions have achieved good performance. The aim of the disentanglement learning in these systems is to remove the speaker information and retain the speech content information.

Most current deep neural network (DNN) based techniques for speech processing tasks such as automatic speech recognition (ASR) \cite{graves2013hybrid,peddinti2015time,povey2016purely} and text-to-speech (TTS) \cite{wang2017tacotron,shen2018natural,ping2017deep} rely on annotated resources or expert knowledge. It still remains a challenge to utilize speech processing techniques for languages that lack resources. Zero Resource Speech Challenge series \cite{versteegh2015zero,versteegh2016zero,dunbar2019zero} aim to explore speech processing techniques for a 'low-resource' situation. In the ZeroSpeech 2015 challenge and ZeroSpeech 2017 challenge, the key objective is to learn a representation of speech which is robust to speaker variations. The ZeroSpeech 2019 challenge was expanded to a multi-task scenario covering both: acoustic unit discovery and voice conversion. Participants were required to submit the obtained acoustic units representation and the converted speech. The ZeroSpeech 2020 challenge consolidates the ZeroSpeech 2017 challenge and the ZeroSpeech 2019 challenge.

In the ZeroSpeech 2015 \& 2017 challenge, the best performance methods were based on Dirichlet Process Gaussian Mixture Model (DPGMM) \cite{chang2013parallel}. As the discriminative acoustic units representation, clustering posterior grams \cite{ansari2017unsupervised,heck2017feature} obtained from the DPGMM have been used. However, these posterior grams are still found to contain the speaker information. In order to learn a speaker-invariant representation, Higuchi et al. \cite{Higuchi2019} used DNN bottleneck feature and speaker adversarial training. Besides, vocal tract length normalization (VTLN) \cite{chen2017multilingual} and fMLLR for speaker normalization \cite{heck2017feature} were also used. Feng et al. \cite{Feng2019} used a disentanglement learning model \cite{hsu2017unsupervised} combined with the DNN bottleneck feature learning system. The state of the art acoustic unit discovery performance \cite{chorowski2019unsupervised} of the ZeroSpeech 2017 challenge was obtained by using VQVAE \cite{van2017neural} with a WaveNet \cite{vanwavenet} decoder and time-jitter regularization.  

More recently, in the context of multi-task learning, the ZeroSpeech 2019 challenge was dominated by use of auto-encoder with discrete latent variable such as VQVAE \cite{van2017neural}
. VQVAE obtains the representation in a discrete latent space and generates the converted speech from a discrete latent space. However, it is notable that VQVAE based systems \cite{Eloff2019,Tjandra2019} suffered from poor quality of the reconstructed speech. We hypothesize that the cause of this phenomenon derives from two main reasons: (1) the speech is generated in a two-stage process with independently trained vocoder models (2) the discrete latent space with low complexity might cause the fine-grained acoustic units information to be lost.   

This work aims to improve the speech quality and retain the discriminability of the representation. In order to learn a better representation of the fine-grained acoustic units, the main idea is to increase the complexity of the latent representation in a WaveNet \cite{van2017neural}  auto-encoder model. The contribution can be summarised as following: (1) an auto-encoder with a WaveNet decoder is used to directly generate waveform data; (2) instead of using vector quantization as in VQVAE, we propose to utilize two alternative disentanglement learning methods. More specifically, we propose to use sliced vector quantization module \cite{kaiser2018fast}, aiming to increase the complexity of the discrete latent space. Furthermore, we propose to add instance normalization \cite{ulyanov2016instance} layers to the WaveNet auto-encoder model, which has a continuous latent representation. 

\section{Proposed Methods}
In this section, methods submitted to the 2019 part of the ZeroSpeech 2020 challenge are introduced. The 2019 part is a multi-task scenario including two sub-tasks: acoustic unit discovery and voice conversion. In the context of multi-task learning, the challenge requires the participants to develop a representation of acoustic units without supervision. The representation are supposed to be speaker-invariant. Based on the representation, the participants need to generate the converted speech given the target speaker ID.

Our submission utilizes two different methods which we name as instance normalization WaveNet auto-encoder (IN-WAE) and sliced vector quantized WaveNet auto-encoder (SVQ-WAE) respectively. These are independent submissions in this challenge. 
IN-WAE is a WaveNet auto-encoder model incorporated with the instance normalization \cite{ulyanov2016instance} layers. SVQ-WAE is a WaveNet auto-encoder model with a sliced vector quantization \cite{kaiser2018fast} bottelneck module.
\begin{figure}[t]
  \centering
  \includegraphics[width=\linewidth]{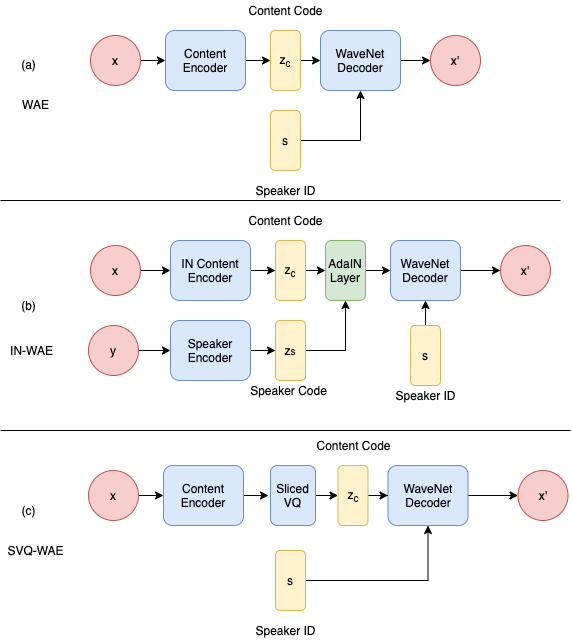}
  \caption{System architectures: (a) WaveNet auto-encoder (WAE), (b) instance normalization WaveNet auto-encoder (IN-WAE), (c) sliced vector quantized WaveNet auto-encoder (SVQ-WAE) $x$ is the input speech, $z_c$ is the content code, $z_s$ is the speaker code, $s$ is the speaker ID input, $x'$ is the reconstructed speech}
  \label{fig:inae}
\end{figure}
\subsection{WaveNet Auto-encoder}
Figure \ref{fig:inae}(a) illustrates the architecture of the WaveNet auto-encoder model \cite{chorowski2019unsupervised}. Let $x \in X$ be the input speech data. The content encoder converts the acoustic unit information into the content code $z_c$. Then the WaveNet \cite{vanwavenet} decoder generates speech data conditioning on the speaker ID $s$ and the content code $z_c$. At training time, $s$ is the source speaker ID, which is associated with the input $x$. The objective function is to reconstruct the input data $x$. At conversion time, $s$ is the target speaker ID, and the model produces converted speech data. As for acoustic unit discovery, the content code $z_c$ is regarded as the representation of acoustic unit information.

The content encoder contains six 1D convolutional layers and four residual ReLU \cite{nair2010rectified} layers. The six convolutional layers can be separated to three groups: (1) 2 layers with kernel size 3 and stride 1; (2) 2 down-sampling layer with kernel size 4 and stride 2; (3) 2 layers with kernel size 3 and stride 1. The down-sampling rate of the content encoder is controlled by the stride of the convolutional layer. For example, two convolutional layers with stride 2 enables a 25 Hz frame rate latent code. The content code $z_c$ and the speaker ID $s$ are the inputs to the WaveNet decoder. The WaveNet model contains 4 up-sampling layers and 20 dilation convolutional layers.

\subsection{Instance Normalization WaveNet Auto-encoder}
Instance Normalization \cite{ulyanov2016instance} (IN) and Adaptive Instance Normalization \cite{huang2017arbitrary} (AdaIN) were proposed for image style transfer. IN normalizes the feature for each sample and each feature channel. Chou et al. \cite{Chou2019} has shown that IN can produce disentanglement between content and speaker information. AdaIN is an extension of IN. AdaIN normalizes the feature as IN, then it adapts the feature to the target style given by a style input data.  

Let $m \in \mathrm{R}^{B \times C \times T}$ be the output feature map of a convolutional layer in a deep neural network, where $B$ is the batch size, $C$ is the number of channels, $T$ is the length of feature frames. Let $m_{b,c,t}$ represents the element of $b$th sample, $c$th channel and $t$th frame. The channel-wise mean $\mu_{c}$ and standard deviation $\sigma_{c}$ of $c$th channel can be obtained by using the following equations.
\begin{equation}
    \mu_{c} = \frac{1}{BT}\sum_{b=1}^{B}\sum_{t=1}^T m_{b,c,t}
\end{equation}
\begin{equation}
    \sigma_{c}^2 = \frac{1}{BT}\sum_{b=1}^{B}\sum_{t=1}^T (m_{b,c,t} - \mu_{c})^2
\end{equation}
The output of the IN layer is the channel-wise normalized feature.
\begin{equation}
    o^{IN}_{b,c,t} = \frac{m_{b,c,t} - \mu_c}{\sqrt{\sigma_{c}^2 + \epsilon}}
\end{equation}
where $o^{IN}_{b,c,t}$ is the normalized feature, $\epsilon$ is a parameter that avoids numerical  instability.

AdaIN receives the style input $y$ and adapts the normalized feature.
\begin{equation}\label{eq:adain}
    o^{AdaIN}_{b,c,t} = \sigma(y) o^{IN}_{b,c,t} + \mu(y)
\end{equation}
where $\sigma(y)$ and $\mu(y)$ are trainable functions.
Following previous work \cite{Chou2019} adding IN and AdaIN layers to the auto-encoder model, the IN-WAE extends the WAE by adding the IN layer to the content encoder and adding the AdaIN layer ahead of the WaveNet decoder. Incorporating with the IN layer, \cite{Chou2019} has shown the content encoder can normalize the global speaker variation while retaining the fine-grained acoustic units information. 
The IN-WAE utilizes the same content encoder architecture as the WAE. The IN layer is added \cite{Chou2019} to the IN content encoder behind every two layers (2,4,6,8,10 th layer).

As shown in Figure \ref{fig:inae}(b), since the AdaIN layer requires a speaker code $z_s$, a speaker encoder is used to derive the speaker information from the speech input $y$. The speaker encoder contains 3 convolutional layers. The global average pooling layer is used as the last layer of the speaker encoder, which compresses the feature into one vector. The output of the speaker encoder is the speaker code $z_s$, which is the input to the AdaIN layers. As in Equation \ref{eq:adain}, $z_c$ is adapted according to the speaker code $z_s$. 
The WaveNet decoder generates speech data conditioning on the output of the AdaIN layer and the speaker ID.
\subsection{Sliced Vector Quantization WaveNet Auto-encoder}
VQVAE \cite{van2017neural} is a variant of variational auto-encoder (VAE) \cite{kingma2013auto}. VQVAE encodes data to a discrete latent space through Vector Quantization (VQ) bottleneck module. 
VQVAE consists of three modules: the encoder, the VQ module and the decoder. The VQ module contains a codebook $M \in \mathrm{R}^{K \times D}$, which is regarded as a collection of $D$ dimensional embeddings, where $K$ is the number of embeddings. $m_k$ is the $k$th embedding in the codebook. Let the output of the encoder be $z_e$, which is the input to the VQ module.
The output of the VQ module $z_q$ is computed as the nearest neighbour of $z_e$ in the latent embedding space $M$.
\begin{equation}
    z_q = argmin_{m_k \in M} ||z_e - m_k||_2
\end{equation}
The decoder receives $z_q$ and reconstructs the data. The objective function of VQVAE can be written as following formula.
\begin{equation}\label{eq:vqvae}
    \mathcal{L} = logp(x|z_q) - ||sg(z_e) - z_q||_2 - \beta ||z_e - sg(z_q)||_2
\end{equation}
The 2nd and 3rd term in Equation \ref{eq:vqvae} are VQ losses, $sg$ is the stop gradient function where the backward gradient is 0. Since the training speed of the encoder and the decoder are different, $sg$ function is used to learn the encoder and the decoder parameters separately. $\beta$ is a hyper-parameter that balances two VQ losses. 
\begin{equation}
    sg(x) = \begin{cases}
            x & forward\\
            0 & backward
            \end{cases}
\end{equation}
Since the VQ module utilizes $argmin$ function, which is non-differentiable, the straight-through \cite{bengio2013estimating} trick is used for gradient estimation. The straight-through trick maps the gradient from $z_q$ to $z_e$.  

The sliced Vector Quantization (Sliced-VQ) module splits the output of the encoder $z_e$ into $N$ slices. 
\begin{equation}
    z_e = concat(z_e^1,...,z_e^n,...,z_e^N)
\end{equation}
where $z_e^n \in \mathrm{R}^{D // N}$, $concat$ is the concatenate function which concatenates all the feature slices. 
The Sliced-VQ operates $N$ parallel sub-VQs on the feature slices $\{z_e^n\}_{1}^N$. The sub-codebook for each sub-VQs were defined as $\{M^n\}_{1}^N$ where $M^n \in \mathrm{R}^{K \times (D // N)}$. The output of each sub-VQ module can be computed as:
\begin{equation}
    z_q^n = argmin_{m_k^n} ||z_e^n - m^n_k||_2
\end{equation}
Then the concatenation of all $z_q^n$ forms the final output of the Sliced-VQ module.
\begin{equation}
    z_q = concat(z_q^1,...,z_q^n,...,z_q^N)
\end{equation}

As illustrated in Figure \ref{fig:inae}(c), the content encoder feeds the feature to the Sliced-VQ module, then the output of the Sliced-VQ module is the content code $z_c$ which can be used as the input to the acoustic unit discovery task. The WaveNet decoder receives the content code $z_c$ and speaker ID $s$, then generates the speech data.
\section{Experiment Setup}
The following describes the experiment setup for the 2019 part of the ZeroSpeech 2020 challenge. First, the dataset and the evaluation metrics are introduced. Then the implementation details of the submissions are described.

\subsection{Dataset}
The 2019 part of the ZeroSpeech 2020 challenge corpus contains two languages: English dataset for development and a surprise language \cite{sakti2008development,sakti2008development1} dataset for test. For each language, the dataset is split into four parts: the train unit dataset, the train voice dataset, the train parallel voice set and the test set. The train unit dataset is used for developing acoustic units representations. The English training unit dataset contains 15 hours data for about 100 speakers. The train voice dataset is the training data for target speakers for voice conversion task. The English train voice dataset contains two speakers, 2 hours data per speaker. For surprise language, there is only one target speaker and 1.5 hours data. For the surprise language dataset, the train unit dataset contains 15 hours for 150 speakers and the train voice dataset contains 1 speaker with 1.5 hours data.

In our experiment, the train unit dataset and the train voice dataset are used. The English dataset is used for training and tuning hyper-parameters. The hyper-parameters are kept fixed for training on surprise language dataset. The official evaluation process includes subjective and objective evaluations. The objective evaluation includes two metrics: Machine ABX \cite{schatz2013evaluating} and bit-rate. Both of two objective evaluation metrics focus on the quality of the acoustic unit representation. The machine ABX measures the discriminability of the representations. The bit-rate measures the compression rate of the representation. The subjective evaluation includes: mean opinion score (MOS), similarity and character error rate (CER). Both the MOS score and the similarity are a scalar in range [1,5]. The MOS score represents the naturalness of the converted speech, while the similarity represents the similarity with the target speaker. The challenge organisers also conduct human evaluation according to the transcriptions and uses CER to measure the intelligibility of the produced speech. 
\subsection{Implementation}
13-dimensional MFCCs with 10 ms step size and 25 ms window size were used as the speech feature. The MFCCs are concatenated with the first and the second derivatives. Mean and variance normalization is conducted. The length of the speech segment is 32 frames (320 ms) and the output waveform length is 5120 samples. The implementation \footnote{code: \url{https://github.com/MingjieChen/wavenet_autoencoders}} uses the PyTorch \cite{NEURIPS2019_9015} toolkit. The Adam \cite{kingma2014adam} optimizer with learning rate 4e-4 was used. One single GTX-1080Ti GPU is used for training. The batch size is 10. The IN-WAE model is trained for 600k steps on the English training dataset. The SVQ-WAE model is trained for 400k steps . The training takes one day for every 100k steps.
\section{Results}
\subsection{Latent Representation Frame Rate}
\begin{table}[t]
    \centering
    \caption{Hyper-parameter exploration: latent representation frame rate for the IN-WAE, the VQ-WAE and the SVQ-WAE}
    \begin{tabular}{*5c}
    \hline
      {} & \multicolumn{2}{c}{50 Hz} &\multicolumn{2}{c}{25 Hz}\\
      \hline
      Model & ABX & Bit-rate & ABX & Bit-rate \\
      \hline
        IN-WAE  &19.13 & 820.08 & 20.19& 385.75\\
        VQ-WAE & 33.31& 328.50& 31.44 & 163.89\\ 
        SVQ-WAE(2 slices) & 32.68& 587.17 & 26.24 & 376.82\\
        SVQ-WAE(4 slices) & 31.39 & 790.68 & \textbf{26.06} & 377.05\\
        \hline
    \end{tabular}
    
    \label{tab:hyper1}
\end{table}
In Table \ref{tab:hyper1}, the effect of latent representation frame rate on the discriminability and the compression rate of the acoustic units representation is explored. For the IN-WAE, the 50 Hz model obtains better ABX score (19.13) than the model with 25 Hz frame rate (20.19). However, the model with 50 Hz frame rate gets higher bit-rate (820.08) than 25 Hz (385.75). The number of the embeddings in the codebook of the SVQ-WAE model is kept as 128. Table \ref{tab:hyper1} compares the VQ-WAE model, the 2 slices SVQ-WAE and the 4 slices SVQ-WAE. Comparing two frame rate options, the trend is that, 25 Hz frame rate obtains better ABX score than 50 Hz. The best ABX score (26.06) is obtained when frame rate is 25 Hz with 4 slices. 

For the IN-WAE, the 50 Hz model shows better discriminability than the 25 Hz model, because higher compression rate might causes the fine-grained acoustic units information such as short phones to be lost. As for the SVQ-WAE, the 25 Hz model shows better discriminability than the 50 Hz model. Comparing two proposed methods, the IN-WAE shows better discriminability than the SVQ-WAE.
\subsection{Bottleneck Shape in the SVQ-WAE}
The hyper-parameters in terms of the shape of the sliced-VQ module in the SVQ-WAE are explored. This part of experiment explores the number of the embeddings $k$ in codebook of the sliced-VQ module and the number of slices $N$. The frame rate is kept as 25 Hz.
\begin{table}[t]
    \centering
    \caption{Hyper-parameter Exploration for the SVQ-WAE}
    \begin{tabular}{c|c|c|c}
    \hline
    \multirow{2}{*}{$\#$Slices}{} & \multicolumn{3}{c}{ABX/Bit-rate} \\ \cline{2-4}
     {} &  k=128& k=512 &k=1024  \\ 
     \hline
     N=1 & 31.44/163.89&30.91/171.98 &28.05/175.84\\
     N=2 & 26.24/ 376.82& 28.80/339.55& 27.59/303.69 \\
     N=4 & \textbf{26.06}/377.05&26.92/379.52 &27.06/379.76\\
     \hline
    \end{tabular}
    \label{tab:hyper2}
\end{table}
In Table \ref{tab:hyper2}, 
the best ABX score (26.06) is when k is 128 and N is 4, and the best bit-rate (163.89) is when N is 1 and k is 128. The trend is that the ABX score gets lower and the bit-rate gets higher as the number of the slices $N$ increasing. It means that the latent space is getting more complex as $N$ increases. And a higher complexity of the latent space benefits the modeling of the acoustic units information. Moreover, a higher $N$ also means a higher bit-rate. It is also notable that increasing the number of embeddings $k$ has a negative effect on both the ABX score and the bit-rate.
\subsection{ZeroSpeech Challenge Results}
The challenge official baseline method is a combination of a DPGMM \cite{chang2013parallel} system and a Merlin \cite{wu2016merlin} system. The topline method is a combination of an ASR and a TTS systems trained with annotated data.
\begin{table}[t]
    \caption{Comparing results of test dataset (surprise language)}
    \centering
    \begin{tabular}{cccccc}
    \hline
    Model&MOS&CER&Similarity&ABX&Bit-rate  \\
    \hline
    Baseline & 2.23 & 0.67 & 3.26 & 27.46 & 74.55\\
    Topline & 3.49 & 0.33 & 3.77 & 16.09 & 35.2\\
    IN-WAE  & \textbf{4.06} & \textbf{0.15} & 2.67 & \textbf{12.5} & 387.83 \\
    SVQ-WAE & 2.28 & 0.55 & 2.5 & \textbf{16.47} & 384.23\\
    \hline
    \end{tabular}
    \label{tab:submit_result1}
\end{table}
\begin{table}[t]
    \caption{Comparing results of developing dataset (English)}
    \centering
    \begin{tabular}{cccccc}
    \hline
    Model&MOS&CER&Similarity&ABX&Bit-rate  \\
    \hline
    Baseline & 2.14 & 0.77 & 2.98 & 35/63 & 71.98\\
    Topline & 2.52 & 0.43 & 3.1 & 29.85 & 37.73\\
    IN-WAE  & \textbf{3.61} & \textbf{0.18} & 2.57 & \textbf{20.19} & 385.75 \\
    SVQ-WAE & 2.88 & 0.47 & 2.35 & \textbf{26.06} & 377.05\\
    \hline
    \end{tabular}
    \label{tab:submit_result2}
\end{table}

Table \ref{tab:submit_result1} and \ref{tab:submit_result2} present the result of our submissions and the provided official systems on developing dataset (English) and test dataset (surprise language) respectively. The total number of submissions was 22, including two official systems. On the surprise language dataset, the IN-WAE obtains the top position for both the naturalness (MOS score 4.06) and the intelligibility (CER 0.15). Meanwhile, it achieves the third position for discriminability of the representation (ABX score 12.5). However, it is obvious that the speaker similarity is worse than the official baseline system. The SVQ-WAE obtains the fifth position for ABX score, whereas it does not show an advantage on any other metrics. Comparing the IN-WAE and the SVQ-WAE, the IN-WAE with continuous representation has advantages on both speech quality and representation discriminability.
For the English dataset, as shown in Table \ref{tab:submit_result2}, the IN-WAE achieves the third position for both naturalness (MOS score 3.61) and representation discriminability (ABX score 20.19 ) and the top position for intelligibility (CER 0.18). The SVQ-WAE achieves the 6th position on ABX score and the 8 th on MOS score. Both the IN-WAE and the SVQ-WAE are not showing competitive performance for speaker similarity. The reason might be that the latent representation still contains speaker information with an increased complexity. 
\subsection{Challenge Result Comparison}
\begin{figure}[t]
    \centering
    \includegraphics[width=\linewidth]{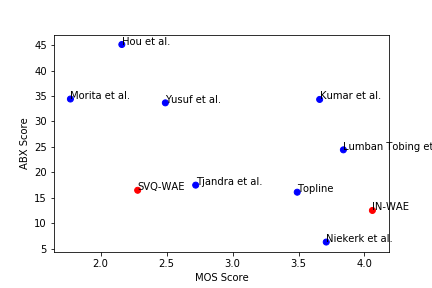}
    \caption{The result comparison of ABX score and MOS score for part of the submissions in ZeroSpeech challenge 2020 test data}
    \label{fig:result_compare}
\end{figure}
We aimed to improve the quality of the speech and meanwhile keep the discriminability of the representation. Figure \ref{fig:result_compare} plots the results of a part of submissions in this challenge. As shown in Figure \ref{fig:result_compare}, the IN-WAE obtains a better MOS score while a worse ABX score than \cite{Niekerk2020}. The SVQ-WAE does not obtain good MOS score however it still gets competitive performance on ABX score.
\section{Conclusions and Future Work}
We proposed to incorporate WaveNet auto-encoder with instance normalization and sliced-VQ respectively. In the ZeroSpeech 2020 challenge, the IN-WAE obtains competitive performance on naturalness, intelligibility and representation discriminability. Meanwhile, the SVQ-WAE obtains competitive representation discriminability. In future work, the methods that can achieve good speaker similarity for voice conversion will be explored. Moreover, the techniques that can accelerate the WaveNet auto-encoder model inference will be investigated.
\bibliographystyle{IEEEtran}

\bibliography{main}


\end{document}